\documentstyle{mn}
\input epsf

\begin{document}

\journal{Preprint astro-ph}

\title{\bf Bumpy Power Spectra and $\mathbf{\Delta T/T}$}

\author[Louise~M.~Griffiths, Joseph~Silk and Saleem~Zaroubi]{Louise M. Griffiths$^1$, Joseph Silk$^1$ and Saleem Zaroubi$^2$\\
$^1$Astrophysics, Nuclear and Astrophysics Laboratory, Keble Road, Oxford
OX1 3RH, United Kingdom.\\
$^2$Max Planck Institute for Astrophysics, Karl Schwarzschild Str. 1, 85748 Garching, Germany.}

\maketitle


\begin{abstract}
With the recent publication of the measurements of the radiation angular power spectrum from the BOOMERanG Antarctic flight (de Bernardis et al. 2000), it has become apparent that the currently favoured spatially-flat cold dark matter model (matter density parameter $\Omega_{\rm m}=0.3$, flatness being restored by a cosmological constant $\Omega_{\Lambda}=0.7$, Hubble parameter $h=0.65$, baryon density parameter $\Omega_{\rm b}h^2=0.02$) no longer provides a good fit to the data.   We describe a phenomenological approach to resurrecting this paradigm.  We consider a primordial power spectrum which incorporates a bump, arbitrarily placed at $k_{\rm b}$, and characterized by a Gaussian in log $k$ of standard deviation $\sigma_{\rm b}$ and amplitude ${\rm A}_{\rm b}$, that is superimposed onto a scale-invariant power spectrum.  We generate a range of theoretical models that include a bump at scales consistent with cosmic microwave background and large-scale structure observations, and perform a simple $\chi^2$ test to compare our models with the $COBE$ DMR data and the recently published BOOMERanG and MAXIMA data.  Unlike models that include a high baryon content, our models predict a low third acoustic peak.  We find that low $\ell$ observations (20 $< \ell <$ 200) are a critical discriminant of the bumps because the transfer function has a sharp cutoff on the high $\ell$ side of the first acoustic peak.  Current galaxy redshift survey data suggests that excess power is required at a scale around 100 Mpc, corresponding to $k_{\rm b}$ $\sim$ 0.05 $h$ Mpc$^{-1}$.  For the concordance model, use of a bump-like feature to account for this excess is not consistent with the constraints made from recent CMB data.  We note that models with an appropriately chosen break in the power spectrum provide an alternative model that can give distortions similar to those reported in the APM survey as well as consistency with the CMB data (Atrio-Barandela et al. 2000; Barriga et al. 2000).  We prefer however to discount the APM data in favour of the less biased decorrelated linear power spectrum recently constructed from the PSCz redshift survey (Hamilton \& Tegmark 2000).  We show that the concordance cosmology can be resurrected using our phenomenological approach and our best-fitting model is in agreement with the PSCz observations.
\end{abstract}


\section{Introduction}
The recent BOOMERanG and MAXIMA measurements of an acoustic peak in the angular power spectrum of the cosmic microwave background (CMB) temperature at $l \approx 200$ (de Bernardis et al. 2000; Hanany et al. 2000) has provided remarkable confirmation that the growth via gravitational instability of primordial adiabatic density fluctuations seeds large-scale structure. One consequence of the location of this peak, due to the compression of an acoustic wave on first entering the horizon of last scattering, is that the spatial geometry of the universe is flat.

However, the weakness of the second acoustic peak at $l \approx 400$, due to the subsequent first rarefaction of the acoustic wave on the last scattering horizon, has provoked considerable speculation as to the additional freedom that could be added to the concordance cold dark matter (CDM) model (matter density parameter $\Omega_{\rm m}=0.3$, flatness being restored by a cosmological constant $\Omega_{\Lambda}=0.7$, Hubble parameter $h=0.65$, baryon density parameter $\Omega_{\rm b}h^2=0.02$) to accommodate such an effect.  Ideas that have been proposed include enhancement of the baryon fraction (Lange et al. 2000; White et al. 2000), a large neutrino asymmetry (Lesgourgues \& Peloso 2000), delay of recombination (Hu \& Peebles 2000), an admixture of a component of cosmological defects (Bouchet et al. 2000) and models employing double inflation in supergravity (Kanazawa et al. 2000).

Here we suggest a more phenomenological solution, which is motivated by suggestive, although not overwhelming, evidence from galaxy surveys that there is excess power relative to the scale-invariant $(n \approx 1)$ fluctuation spectrum of the conventional model near 100 $h^{-1}$ Mpc. The case for excess power has not hitherto been completely convincing because one is probing the limit of current surveys. Nevertheless, several independent data sets have provided such indications (See e.g. Broadhurst et al. 1990; Einasto et al. 1997; Landy et al. 1996).  

In fact the multiple inflationary model of Adams, Ross \& Sarkar (1997) predicts the suppression of the second acoustic peak through the generation of features in the primordial power spectrum from phase transitions that occur during inflation.  More generally, there are strong theoretical arguments which suggest that arbitrary features can be dialled onto the primordial power spectrum predicted by generic inflationary models (See e.g. Chung et al. 1999; Garc\'{\i}a-Bellido et al 1996; Lesgourges et al. 1998; Linde \& Mukhanov 1997; Martin et al. 1999; Randall et al 1996; Starobinsky 1998).  

We therefore consider a primordial power spectrum which incorporates a phenomenological bump, arbitrarily placed at $k_{\rm b}$, and characterized by a Gaussian in log $k$ of standard deviation $\sigma_{\rm b}$ and amplitude ${\rm A}_{\rm b}$, that is superimposed onto a scale-invariant power spectrum as advocated by Silk \& Gawiser (1998). We examine the constraint on the bump parameters for the $\Lambda$CDM concordance model (Ostriker \& Steinhardt 1995) ($\Omega_{\rm m}=0.3, \, \Omega_{\Lambda}=0.7, \, h=0.65, \, \Omega_{\rm b}h^2=0.02)$ posed by the CMB data, and restrict the choice of bump parameters to the region of parameter space that is consistent with observations of large-scale power and CMB anisotropies.

\section{The Theoretical Models}
In our paper we consider one particular spatially-flat CDM model; the $\Lambda$CDM concordance model of Ostriker \& Steinhardt (1995) ($\Omega_{\rm m}=0.3, \, \Omega_{\Lambda}=0.7, \, h=0.65, \, \Omega_{\rm b}h^2=0.02)$.  We assume Gaussian and adiabatic initial conditions with a scale-invariant ($n=1$) power law form as predicted by the simplest inflationary models.  The radiation angular power spectrum is calculated using the {\sc cmbfast} program (Seljak \& Zaldarriaga 1996) once the code has been modified to incorporate a bump in the primordial spectrum as advocated by Silk \& Gawiser (1998).  We model this bump as a Gaussian in log $k$ with a central location in wavenumber $k_{\rm b}$, a standard deviation $\sigma_{\rm b}$, and an amplitude A$_{\rm b}$, resulting in the new primordial power spectrum given below, where $P_{0}(k)$ is the power spectrum of the model without the feature.

\begin{equation}
P(k) \, = \, P_0(k) \, \left(1 \, + \, {\rm A}_{\rm b} \, \exp\left(- \frac{(\log \, k \, - \, \log \, k_{\rm b})^2}{2 \, \sigma_{\rm b}^2}\right)\right) \,,
\end{equation}

We restrict the choice of bump parameters to the region of parameter space that is consistent with large-scale structure and CMB observations.  The parameters are varied as follows: $0.05 \, < \, \sigma_{\rm b} \, < \, 2.0$, $0.0 \, < \, {\rm A}_{\rm b} \, < \, 3.0$, $0.001 \, < \, k_{\rm b} \, h \, {\rm Mpc}^{-1} \, < \, 0.140$.

Our focus is directed towards determining whether it is possible to resurrect the concordance model without resorting to the proposed ideas listed in our introduction that may prove to contradict observation.  Since the CMB observations indicate that the second acoustic peak is suppressed in relation to the first acoustic peak, it may be that a dip in the primordial power spectrum around the scale of the second peak could also enable the concordance model to fit the data.  Theories that predict a bump in the primordial power spectrum have been inspired by hints of such a feature from large-scale structure observations.  We do not investigate a dip in this paper because there is less theoretical motivation for this scenario.  We attempt to increase the first to second peak ratio with the incorporation of a bump around the scale of the first peak, then renormalize the radiation angular power spectrum to fit the data.

\begin{table}
\begin{center}
\caption{\label{obsdat} The data used in this study.}
\vspace{2pt}
\begin{tabular}{|l|l|r|c|} \hline         
Experiment & $\ell_{\rm eff}$ & $\delta T_{\ell_{{\rm eff}}}^{{\rm data}} \pm \sigma^{{\rm data}}(\mu$K$^2)$\\
\hline
COBE &2.1&$ 72.25_{-72.5}^{+528.0}$\\
COBE &3.1&$ 784.0_{-470.71}^{+476.25}$\\
COBE &4.1&$ 1156.0_{-437.76}^{+444.0}$\\
COBE &5.6&$ 630.01_{-287.76}^{+294.15}$\\
COBE &8  &$ 864.36_{-224.27}^{+224.64}$\\
COBE &10.9&$ 767.29_{-229.05}^{+231.27}$\\
COBE &14.3&$ 681.21_{-244.4}^{+249.04}$\\
COBE &19.4&$ 1089.0_{-327.24}^{+324.76}$\\
BOOMERanG &50.5&$ 1140 \pm280$\\
BOOMERanG &100.5&$ 3110 \pm490$\\
BOOMERanG &150.5&$ 4160 \pm540$\\
BOOMERanG &200.5&$ 4700\pm540$\\
BOOMERanG &250.5&$ 4300\pm460$\\
BOOMERanG &300.5&$ 2640\pm310$\\
BOOMERanG &350.5&$ 1550\pm220$\\
BOOMERanG &400.5&$ 1310\pm220$\\
BOOMERanG &450.5&$ 1360\pm250$\\
BOOMERanG &500.5&$ 1440\pm290$\\
BOOMERanG &550.5&$ 1750\pm370$\\
BOOMERanG &600.5&$ 1540\pm430$\\
MAXIMA &73&$ 2000_{-510}^{+680}$\\
MAXIMA &148&$ 2960_{-550}^{+680}$\\
MAXIMA &223&$ 6070_{-900}^{+1040}$\\
MAXIMA &298&$ 3720_{-540}^{+620}$\\
MAXIMA &373&$ 2270_{-340}^{+390}$\\
MAXIMA &448&$ 1530_{-270}^{+310}$\\
MAXIMA &523&$ 2340_{-380}^{+430}$\\
MAXIMA &598&$ 1530_{-340}^{+380}$\\
MAXIMA &673&$ 1830_{-440}^{+490}$\\
MAXIMA &748&$ 2180_{-620}^{+700}$\\
\hline
\end{tabular}
\end{center}
\end{table}

\section{The Observational Data}
Our data sample is listed in Table 1.  It consists of the 8 uncorrelated $COBE$ DMR points from Tegmark \& Hamilton (1997), the 12 data points from the BOOMERanG Antarctic flight (de Bernardis et al. 2000) and the 10 recently published MAXIMA data points (Hanany et al. 2000).

\begin{figure}
\centering 
\leavevmode\epsfysize=7cm\epsfbox{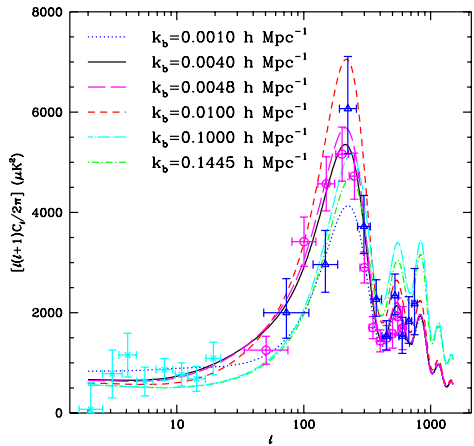}\\ 
\caption[k_bmp]{\label{f:k_bmp} The observational data set of Table 1. The crosses indicate the eight uncorrelated $COBE$ DMR points (Tegmark \& Hamilton 1997), the circles indicate the twelve BOOMERanG data points (de Bernardis et al. 2000) and the triangles indicate the ten MAXIMA data points (Hanany et al. 2000).  The solid curve shows the best-fitting model ($k_{\rm b} = 0.004 \, h \, {\rm Mpc}^{-1}$, $\sigma_{\rm b} = 1.05$, ${\rm A}_{\rm b} = 0.9$) normalized to the full observational data set.  The remaining curves show the same model with varying $k_{\rm b}$ as indicated.  All models are normalized to the best-fitting model.}
\end{figure}

\section{Constraining the Models}
We use a simple $\chi^2$ goodness-of-fit analysis employing the data in Table 1 along with the corresponding window functions for the uncorrelated $COBE$ DMR points (Tegmark \& Hamilton 1997) and assuming a top hat window function over the BOOMERanG and MAXIMA bins.  The window functions describe how the anisotropies at different $\ell$ contribute to the observed temperature anisotropies (Lineweaver et al. 1997).  For a given theoretical model, they enable us to derive a prediction for the $\delta T$ that each experiment would see, to be compared with the observations in Table 1.

It has been noted that the use of the $\chi^2$ test can give a bias in parameter estimation in favour of permitting a lower power spectrum amplitude because in reality there is a tail to high temperature fluctuations. Other methods have been proposed (Bond, Jaffe \& Knox 1998; Bartlett et al. 1999) which give good approximations to the true likelihood, though they require extra information on each experiment which is not yet readily available. We do not use these more sophisticated techniques here.

There are $N_{{\rm data}} = 30$ data points.  Rather than adopting the $COBE$ normalization, the theoretical models are normalized to the full observational data set resulting in a hidden parameter.  We use the method of Lineweaver \& Barbosa (1998) to treat the correlated calibration uncertainty of the 12 BOOMERanG data points and that of the 10 MAXIMA data points as free parameters with Gaussian distributions about their nominal values of 10\% for BOOMERanG and 4\% for MAXIMA.  This results in two further hidden parameters.  We do not account for the 10\% correlation between the BOOMERanG bins nor that between the MAXIMA bins which would further reduce the degrees of freedom.  Accounting for the correlations would provide tighter constraints on the models, so the constraints we make are conservative.

Because we are measuring absolute goodness-of-fit on a model-by-model basis, with three hidden parameters, the appropriate distribution for the $\chi^2$ statistic has $N_{{\rm data}}$ - 3 degrees of freedom. Nothing further is to be subtracted from this to allow for the number of parameters, as they are not being varied in the fit. To assess whether a model is a good fit to the data, we need the confidence levels of this distribution. These are $\chi^2_{27} <$ 29.87 at the 68\% confidence level, $\chi^2_{27} <$ 40.11 at the 95\% confidence level and $\chi^2_{27} <$ 46.96 at the 99\% confidence level. Models which fail these criteria are rejected at the given level.  

Although we are unable to give the overall best-fitting model for currently permitted cosmologies, since this would require varying each of the cosmological parameters as well as those describing the bump, we find that for the paradigm considered the best-fitting model is $k_{\rm b}=0.004 \, h \, {\rm Mpc}^{-1}$, \, ${\rm A}_{\rm b}=0.9$, \, $\sigma_{\rm b}=1.05$.  This model has a $\chi^{2}$ of 22.0 which is in good agreement with expectations for a fit to 30 data points with 6 adjustable parameters (the 3 bump parameters and the 3 hidden parameters).  

By marginalizing over the bump parameters we are able to determine the 68\% confidence level limits on each parameter.  We find that the lower limit on $k_{\rm b}$ extends right to the edge of the region of parameter space that we are investigating.  We do not feel it necessary to push this limit further since it extends into the region of greatest observational uncertainty due to cosmic variance.   The upper limits in both $\sigma_{\rm b}$ and A$_{\rm b}$ also reach the edges of our parameter space indicating that the CMB data allows a lot of freedom with the amplitude and standard deviation of a bump.  We find that at the 68\% confidence level $k_{\rm b}\leq 0.014 \, h \, {\rm Mpc}^{-1}$, $\sigma_{\rm b} \geq 0.15$ and A$_{\rm b} \geq 0.3$.

\begin{figure}
\centering 
\leavevmode\epsfysize=7cm\epsfbox{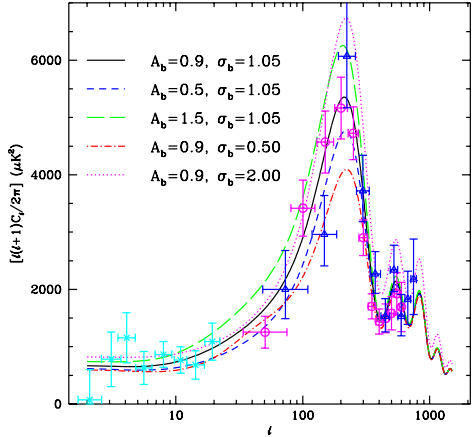}\\ 
\caption[varsiga]{\label{f:varsiga} The same data sample as in Figure~\ref{f:k_bmp}.  The solid curve shows the best-fitting model ($k_{\rm b} = 0.004 \, h \, {\rm Mpc}^{-1}$, $\sigma_{\rm b} = 1.05$, ${\rm A}_{\rm b} = 0.9$) normalized to the full observational data set.  The dotted curve shows the same model with $\sigma_{\rm b}$ = 2.0, the dotted--dashed curve the same model with $\sigma_{\rm b}$ = 0.5 and the dashed curves the same model with ${\rm A}_{\rm b}$ = 0.5 (small dashes) and 1.5.  All models are normalized to the best-fitting model.}
\end{figure}

In Figure~\ref{f:k_bmp} we show the best-fitting model as well as a range of $k_{\rm b}$ models with the same A$_{\rm b}$, $\sigma_{\rm b}$ and normalization to illustrate the effect of varying $k_{\rm b}$.  In Figure~\ref{f:varsiga} we plot the best-fitting model together with models of varying $\sigma_{\rm b}$ and A$_{\rm b}$.  From these figures it can be seen that, unlike models incorporating a high baryon content, our model predicts a low third acoustic peak.  Also these figures highlight that low $\ell$ observations (20 $< \ell <$ 200) are a critical discriminant of the bumps because beyond the first acoustic peak the models become less distinguishable.  This  because the transfer function has a much sharper cutoff on the high $\ell$ side of the first peak, relative to the cutoff on the low $\ell$ side.

\begin{figure}
\centering 
\leavevmode\epsfysize=7cm\epsfbox{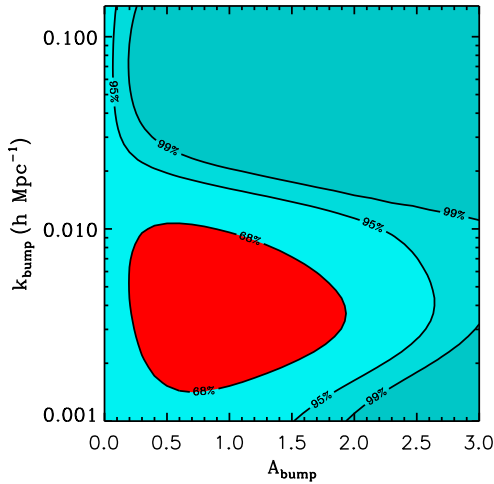}\\ 
\caption[contour]{\label{f:contour} Confidence level contours for the concordance $\Lambda$CDM model as a function of $k_{\rm b}$ and A$_{\rm b}$, $\sigma_{\rm b}$ fixed at 1.05.  The region within the 68\% contour line is allowed at the 68\% confidence level.}
\end{figure}

A confidence level contour map of $k_{\rm b}$ versus A$_{\rm b}$ for the cosmology of interest with $\sigma_{\rm b}$ fixed at 1.05 is shown in Figure~\ref{f:contour}.  This indicates that, for the chosen cosmology, the scale at which a bump can appear in the primordial power spectrum is quite constrained by the CMB data sample.  At the 68\% confidence level we are limited models with a bump at scales $k_{\rm b} \leq 0.010 \, h$ Mpc$^{-1}$ for this value of $\sigma_{\rm b}$, although we have rather more freedom with the amplitude of the bump.

\section{From CMB to Galaxy Surveys}
\begin{figure}
\centering 
\leavevmode\epsfysize=7cm\epsfbox{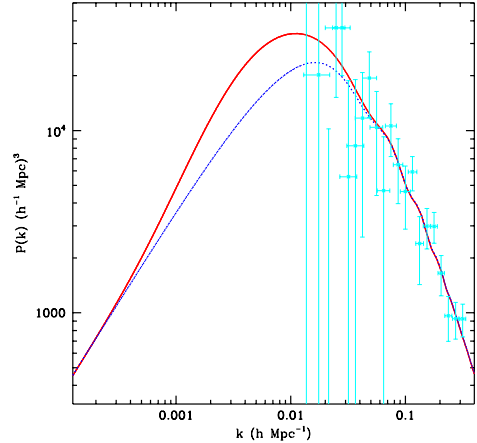}\\ 
\caption[varsiga]{\label{f:density} The PSCz decorrelated linear primordial power spectrum (Hamilton \& Tegmark 2000).  The solid curve shows the standard spatially-flat cosmological model with a bump at $k_{\rm b}=0.004 \, h$ Mpc$^{-1}$, normalized to the CMB data sample with a bias factor of 1.07.  The dotted curve shows the standard model without the bump, normalized to the CMB data sample with a bias factor of 1.16.}\end{figure}
Since our incorporation of a bump into the primordial power spectrum of density perturbations was inspired by large-scale structure observations, it is interesting to ask how our modified model compares with the decorrelated linear power spectrum that was recently generated from the PSCz catalogue (Hamilton \& Tegmark 2000).  We treat the 22 decorrelated PSCz data points as uncorrelated so that the theoretical model that we find to be the best-fit to the CMB observational data set can be compared with the galaxy survey observations using a $\chi^2$ test.   The theoretical model is renormalized to the observational data set resulting in a hidden parameter.  This allows for a bias factor $b$ where
\begin{equation}
P(k)_{\rm \, PSCz} =  b^2 \, P(k)_{\rm \, CMB} \,,
\end{equation}
We note that non-linearity corrections to the data are omitted in our comparison.  In $\Lambda$CDM models the effects of nonlinearity in the matter power spectrum are partially cancelled by galaxy-to-mass antibias, so that the PSCz power spectrum is close to linear all the way to $k = 0.3 \, h$ Mpc$^{-1}$ [Hamilton 2000].

Figure~\ref{f:density} plots our best-fitting CMB normalized standard low-density cosmological model with and without the bump against the PSCz decorrelated linear power spectrum.  Our best-fitting model has been renormalized to the PSCz data set with a bias parameter of 1.07 and the model without the bump takes a bias parameter of 1.16.  Both models are a very good fit to these up-to-date large-scale structure observations ($\chi^2_{\rm best-fit}=17.20$, $\chi^2_{\rm no \, bump}=16.35$, $\chi^2_{21} <$ 23.46 at the 68\% confidence level), but it is clear that the current data does not probe the scales that are critical to discriminating between the models.

\begin{figure}
\centering 
\leavevmode\epsfysize=7cm\epsfbox{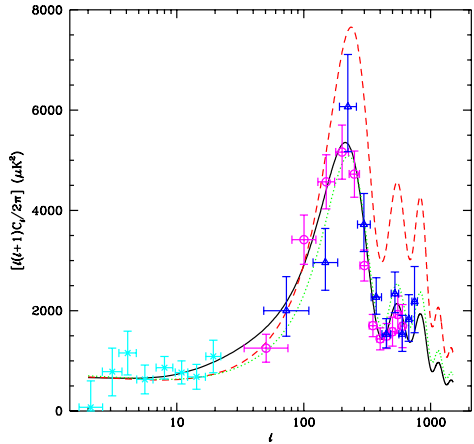}\\ 
\caption[100mpc]{\label{f:100mpc} The same data sample as in Figure~\ref{f:k_bmp}.  The solid curve shows the best-fitting model ($k_{\rm b} = 0.004 \, h \, {\rm Mpc}^{-1}$, $\sigma_{\rm b} = 1.05$, ${\rm A}_{\rm b} = 0.9$), the dotted curve shows the standard $\Lambda$CDM model without a bump and the dashed curve shows the same model with a bump at $k_{\rm b} = 0.052 \, h$ Mpc$^{-1}$, $\sigma_{\rm b} = 1.05$, ${\rm A}_{\rm b} = 0.9$.  Each model is independently normalized to the full CMB observational data set.}\end{figure}

As stated in our introduction, several independent large-scale structure data sets provide suggestive, although not overwhelming, evidence that there is excess power relative to the scale-invariant $(n \approx 1)$ fluctuation spectrum of the conventional model near 100 $h^{-1}$ Mpc, corresponding to $k_{\rm b} \sim 0.05 \, h$ Mpc$^{-1}$ (See e.g. Broadhurst et al. 1990; Einasto et al. 1997; Landy et al. 1996).  Figure~\ref{f:100mpc} shows our best-fitting bump model together with the standard $\Lambda$CDM model without a bump and the same model with a bump at $k_{\rm b} = 0.05 \, h$ Mpc$^{-1}$, each model independently taking its optimal normalization to the full CMB data set.  It is interesting to note that a bump in the primordial power spectrum of any amplitude or standard deviation at $k_{\rm b} = 0.05 \, h$ Mpc$^{-1}$ is ruled out at the 95\% confidence level by the CMB observational data for the $\Lambda$CDM concordance model.  

\section{Summary}
We describe a toy model for a bump to be included in the primordial density power spectrum with the hope of reviving the standard model without resorting to revising fundamental cosmological theories.  We have confronted our theoretical models with the recent BOOMERanG and MAXIMA data and have shown that it is indeed possible to resurrect the standard model, although we are somewhat restricted with where we place our additional feature.  We find that our model, unlike models that include a high baryon content, predicts a low third acoustic peak.

There are two CMB measurements that will help to discriminate between such a bump and cosmological alternatives for suppressing the second peak.  One is of course the detection of the third peak.  In addition, although low $\ell$ (20-100) observations have received relatively little attention and hence are currently a poor constraint on cosmological parameters, we have found that the low $\ell$ power is potentially a critical discriminant for the possible bump feature. This  because the transfer function has a much sharper cutoff on the high $\ell$ side of the first peak, relative to the cutoff on the low $\ell$ side.

Current galaxy redshift survey data suggests that excess power is required at a scale around 100 Mpc, corresponding to $k_{\rm b}$ $\sim$ 0.05 $h$ Mpc$^{-1}$ (See e.g. Broadhurst et al. 1990; Einasto et al. 1997; Landy et al. 1996).  For the concordance paradigm, use of a bump-like feature to account for this excess is not consistent with the constraints from the CMB data.  We note that  models with an appropriately chosen break in the power spectrum provide an alternative model that can give distortions similar to those reported in the APM survey as well as consistency with the CMB data (Atrio-Barandela et al. 2000; Barriga et al. 2000).  We prefer however to discount the APM data in favour of the less biased decorrelated linear power spectrum recently constructed from the PSCz redshift survey (Hamilton \& Tegmark 2000).  

The incorporation of a bump in the primordial spectrum at a scale of $k_{\rm b} = 0.004 \, h$ Mpc$^{-1}$, as the CMB data prefers, is in good agreement with the PSCz power spectrum with a bias parameter of 1.07.  Future surveys such as 2DF and SDSS should be able to probe the large-scale structure power spectrum at the depths required to further test our conjecture. Large-scale velocity field data  are useful only at higher $k$ as a discriminant of bump-like features, and we will address this issue in a later paper.

\section*{ACKNOWLEDGEMENTS}

LMG would like to thank Pedro Ferreira and Andrew Liddle for providing  productive insights into the analysis and understanding of this problem and Alessandro Melchiorri and Bill Ballinger for useful discussions.  We acknowledge use of the Starlink computer system at the University of Oxford and use of the {\sc cmbfast} code of Seljak \& Zaldarriaga (1996).



\bsp

\end{document}